\documentclass[journal=acsnano,manuscript=article]{achemso}
\usepackage[version=3]{mhchem} 
 \usepackage{graphicx} 

\author{James Kestyn}
\author{Sigfrid K. Yngvesson}
\author{Eric Polizzi}
\affiliation{Department of Electrical and Computer Engineering, University of Massachusetts, Amherst}
\email{epolizzi@engin.umass.edu}

\title{One-dimensional Plasmons and Hybridized Coupled Polaritons in Carbon Nanotubes}
\keywords{Tomonaga-Luttinger 1D plasmon, plasmon velocity, 1D polariton, TDDFT, Carbon Nanotubes, High harmonic generation}

\begin{document}

%
%
%
%
%


\begin{abstract}
  This paper presents real-time time-dependent density functional theory (TDDFT) ab-initio
  simulations of selected armchair carbon nanotubes (CNTs). 
  By scaling the lengths of CNTs, 
  we provide a comprehensive analysis of the
  Tomonaga-Luttinger (T-L) 1-D  plasmon velocities, confirming consistency with theoretical predictions and experimental observations.
  Our findings include detailed visual representations of excitation densities at various resonances.
  Furthermore, we explore the coupling between T-L plasmons and
  single electron excitations, identifying distinct 1-D polariton behaviors, such as strong harmonic generation due to nonlinearities, as well as energy gaps that
  differ from conventional 2-D polaritons. The study highlights the unique properties of armchair
  SWCNTs as low-loss nanocavity resonators, demonstrating potential applications in strong light-matter coupling and other nanophotonic devices. 
  The simulation framework employed here opens avenues for further research into 1-D plasmonic phenomena and electronic spectroscopy in complex nanostructures.
\end{abstract}

\section{Introduction}\label{sec1}

The recent surge of activities in Nanoplasmonics\cite{barnes2003,ozbay2006,stockman2011,halas2009,novotny2012}
has shown the potential for realizing functions such as guiding waves
\cite{barnes2003,ozbay2006}, constructing high-Q resonant circuits \cite{wang2020} and antennas \cite{novotny2011}, detecting
and imaging single molecules \cite{cang2011}, greatly enhanced Raman
scattering \cite{zhang2013}, plasmon laser sources \cite{stockman2008}
and many other emerging devices and integrated circuits on a scale much
less than infrared or optical wavelengths. The concept of plasmons was
introduced in the 1950's following the fundamental work by Pines and
Bohm \cite{pines1952,pines1966}. Plasmons were later studied as surface
plasmons \cite{barnes2003} propagating at metal/dielectric interfaces
(2-D), resonating in nanoparticles \cite{scholl2012} (3-D)
 or on graphene (2-D) \cite{garciadeabajo2014}.  Further, as smaller and
smaller 1-D atomic objects were studied it became clear that they had to be
given a full quantum mechanical treatment \cite{tomonaga1950,luttinger1950,deshpande2010,haldane1981,saito1998}. Due
to the very strong electron-electron interactions in 1-D, Landau's Fermi
liquid theory had to be replaced with the Tomonaga-Luttinger
theory in describing the low energy excitations of the electron gas,
so-called Tomonaga-Luttinger (T-L) plasmons. We employ Time
Dependent Density Functional Theory (TDDFT) techniques \cite{marques2012,polizzi2015} to
simulate T-L plasmons in a model 1-D system, armchair metallic
single-walled carbon nanotubes (mSWCNTs). Our recent advances in
numerical scalability and performance of TDDFT simulations (see section on Methods),
have allowed us to efficiently handle systems ranging from molecular
resonances to extrapolations for 1-D metallic wires.  We have also been able to analyze in detail
resonances at higher frequencies than the T-L plasmons, including
excitons and dipolar plasmons \cite{gharbavi2015,spataru2004,hopfield1958}.  Localized dipolar plasmons have resonant energies not as
strongly dependent on L. Similar plasmon resonances were for example
found in simulations of acene molecules up to 7 unit cells
(uce) \cite{guidez2013}. The L-dependence of the dipolar plasmons gives us a handle for adjusting their frequencies such
that they approach the energies of the band-to-band exciton
transitions \cite{gharbavi2015,spataru2004} (which do not vary with L). As L
\emph{increases} the different resonances develop strong and unique
hybridized coupling effects, in distinction from the 2-D CNT film
polaritons discussed in Ref.~\citenum{ho2018,chiu2017,gao2018,haroz2013,forn-diaz2019}
for which the coupling
increases as L \emph{decreases}. The 2-D studies have shown that the
coupling gives rise to avoided resonance crossings and polariton quasi
particle states. In the 1-D case we find that the
results do not fit a standard Rabi coupling model \cite{hopfield1958}. A unique
effect is that additional smaller periodic splittings develop for some
of the resonances. Using targeted TDDFT simulations with sinusoidal
excitation for the (3,3) mSWCNT case we show that the resonances produce harmonic responses when
sinusoidally excited.  The 1-D structures
we have simulated thus offer unique opportunities to study so far
unknown types of strong light-matter coupling phenomena and show the potential for  many novel applications in the area
of quantum electrodynamics.  One dimensional
plasmons have the general advantage of lower propagation losses, strong
spatial confinement and low dispersion compared with surface
polaritons \cite{wang2020,khurgin2015,baranov2018}.

In a previous paper \cite{polizzi2015} we discussed the universal
nature of plasmonic excitations in finite size carbon chains, and two
types of narrow graphene nanoribbons (GNRs), as well as carbon nanotubes.  Kane, Balents and
Fisher \cite{kane1997} showed that the collective boson states in
armchair mSWCNTs lead to the formation of charge density waves (T-L
plasmons) that propagate at a velocity $v_P =v_F/g$ where $v_F$ is the Fermi velocity
($\simeq 10^6 m/s$ \cite{saito1998}) and $g$ is a parameter
that depends on the strength of the electron interactions and the (weak)
screening of the electrostatic environment. For repulsive interactions,
$g<1$ and thus $v_P> v_F$.
It is useful to have a physical picture of the plasmon resonances in
finite size 1-D nano structures by employing an analogy to transmission
line resonances \cite{burke2002,bockrath1999}. The propagating plasmon waves
resonate as they reflect at the ends of a 1-D conductor of length L,
with resonant frequencies given by Eq.~\ref{eq1} (the Plasmon wavelength $\lambda_P$ is 2L).

\begin{equation}
f = \frac{v_P}{2L}
\label{eq1}
\end{equation}

Two Terahertz experiment on a single mSWCNT identified resonances due to
single quasi-particles (g=1)  \cite{zhong2008}, or
g-values from 1 and up to what was expected from the T-L
theory \cite{chudow2012}. Terahertz measurements on mSWCNT
\emph{films} confirmed the rough average value of the Plasmon velocity
in mSWCNTs \cite{zhang2013b,muthee2016}, while recent much more direct
experiments \cite{wang2020,shi2015,wang2019,zhao2018}, employing mid infrared
scattering-type scanning near-field optical microscopy (s-SNOM),
obtained good agreement with the plasmon velocity predicted by Kane et
al. \cite{kane1997} in fairly long armchair single mSWCNTs with half
plasmon wavelengths in the range 35-48nm. Similar s-SNOM measurements of T-L plasmons
were reported in Ref.~\citenum{tian2021} . Ref.~\citenum{tian2018} performed s-SNOM
experiments on \emph{chemically} \emph{doped} semiconducting SWCNTs
. We have simulated (3,3) and (7,7) mSWCNTs for lengths from
1.25 nm to 20 nm in the (3,3) case, an order-of-magnitude longer than in
the earlier paper \cite{polizzi2015}, and from 1.25 nm to 5 nm for the
(7,7) case. We predict plasmon velocities for these lengths that are
less than what the theory in Ref.~\citenum{kane1997} predicts.
Extrapolation of the simulated plasmon velocities are consistent with
that theory for longer lengths in the (3,3) case, however.

Other recent literature has discussed the relationship between Plasmon
resonances and single electron (exciton) transitions, such as for
example Bernadotte et al. \cite{bernadotte2013} . In that paper TDDFT
simulations of molecules, including some 1-D cases, were performed while
varying the electron-electron interaction term. Plasmon resonances
depend strongly on this term, but single electron transitions do not,
providing a method for distinguishing the two types.  Later, the introduction of a generalized plasmonicity index
(GPI) \cite{zhang2017,molinari2015} has helped to further clarify the
distinction between plasmon and single electron resonances. We have
characterized some of the resonances found in this paper in terms of
their GPI. The GPI value expresses the degree to which a resonance
produces an enhancement of the electric field compared with the incident
field, a general characteristic of plasmons. We also present 1-D plots
and 4-D plots of the excited electron density frozen in time and these
plots yield even more detailed information about the nature of the
different resonances.
.

\section{Results and discussion}

\begin{figure}[htbp]
    \centering
    \includegraphics[width=0.75\linewidth]{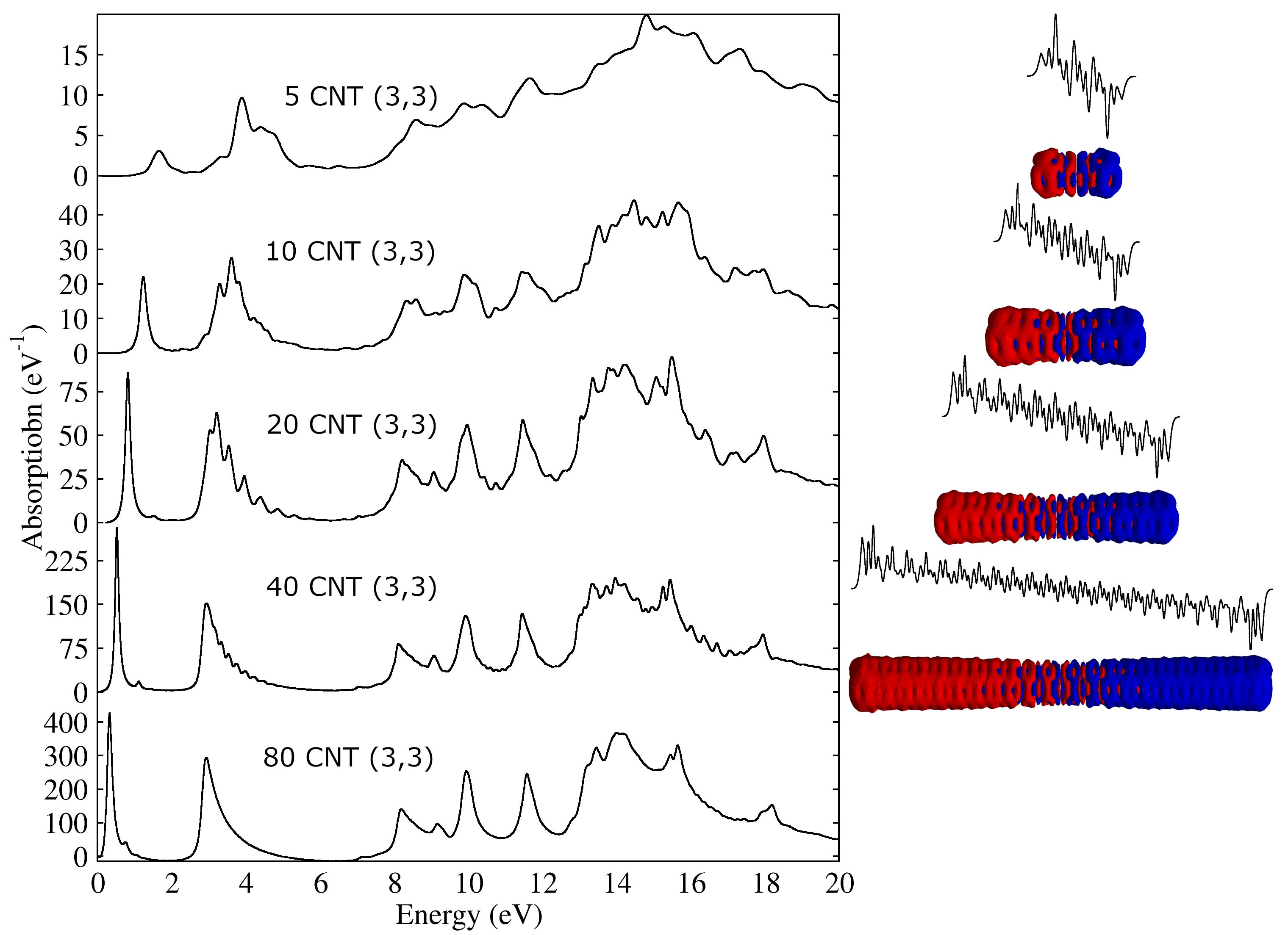}
    \caption{Spectra of the (3,3)mSWCNT for z-directed excitation for lengths of 5, 10, 20, 40 and 80 uce. Also shown are 4-D and 1-D plots of
the excitation densities for the lowest energy peak.}
    \label{fig1}
\end{figure}

\begin{figure}[htbp]
    \centering
    \includegraphics[width=\linewidth]{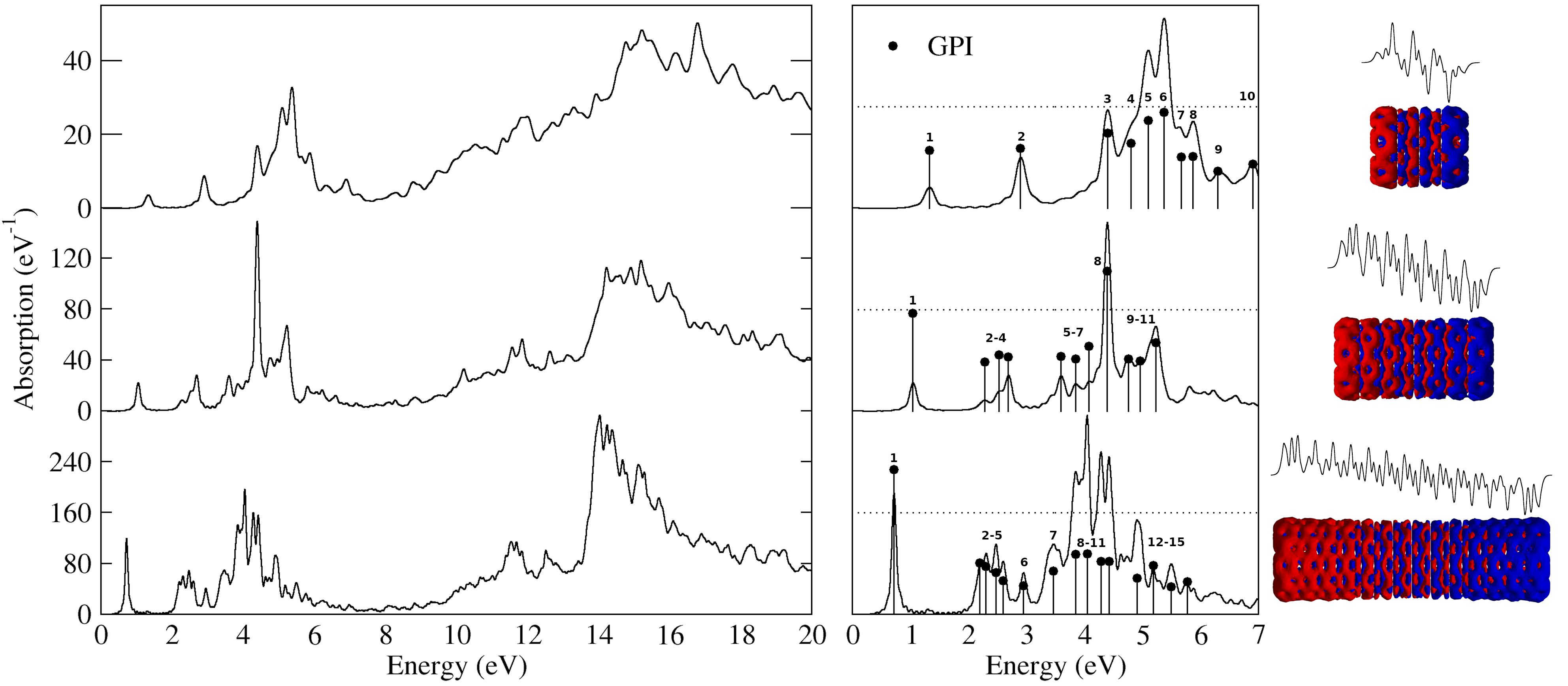}
    \caption{Spectra of 5, 10 and 20 unit cell (7,7) mSWCNTs for
z-direction excitation. GPI is also presented for select peaks, with
values exceeding unity (i.e., the dotted line) defined as plasmon
resonances as opposed to single particle excitations. Response densities
of the lowest energy peaks are shown in the inset (4-D plots and 1-D
density after integration along the cross sections of the tube). Peak
numbers are referred to in the text.}
    \label{fig2}
\end{figure}

Our simulations of 5, 10, 20, 40, and 80 unit cell (uce) (3,3) mSWCNTs
and 5, 10, and 20 uce (7,7) mSWCNTs show evidence of both plasmon and
single particle band-to-band (exciton) transitions, as well as
hybridized combinations of these. The frequency dependent charge density
is expected to oscillate back and forth between the ends of the 1-D
structure and vary across the length of the molecule for a strong
plasmonic resonance with a positive maximum at one end (red in the 4-D plots) and a negative
maximum (blue in the 4-D plots) at the other end. Single electron (band-to-band) transitions,
however, should show localized charge oscillations and densities that
remain relatively flat longitudinally. These characteristics can be seen
in the plots in Figures~\ref{fig1} and \ref{fig2} (for the lowest energy (plasmon) peaks)
as well as in the following sections, where isosurfaces give a snapshot
of the electron response density at that frequency, and 1-dimensional
plots show the cumulative integrated response density over the
cross-section at points along the tubes.

As mentioned in the introduction, recent literature has attempted to
quantify the relationship between the two types of resonances with the
concept of a generalized plasmonicity index (GPI). Interesting
resonances in the (7,7) mSWCNT spectrum have been characterized in terms
of their GPI values. 
We identify the lowest frequency resonances in these spectra
as plasmons (or incipient plasmons) with plasmon velocities (Eq.~\ref{eq1})
that extrapolate toward values that match the predictions of the T-L
theory \cite{kane1997} for tubes about 40 nm length (160uce). The
experiments in Ref.~\citenum{shi2015,wang2019,zhao2018} used tubes in this length range
and demonstrated agreement with the T-L plasmon theory of
Ref.~\citenum{kane1997}. This will be discussed further below.

These characteristics, however, can become unclear due to the presence
of a third possible type of resonance corresponding to hybridized
polariton states \cite{ho2018,chiu2017,gao2018,haroz2013,forn-diaz2019,kockum2018,baranov2018}, with intermediate values for the GPI
and a mixed representation of the charge density. This situation is not
extensively covered in the literature on TDDFT simulations of 1-D nano
objects but was earlier discussed by Bernadotte et al. \cite{bernadotte2013} who
mentioned that, for a certain wave vector range, the plasmonic
excitations will mix with the single-particle transitions. This leads to
a break-down of the plasmon/single-particle picture as they can no
longer be treated as separate excitations. Ref.~\citenum{bernadotte2013} simulated
the effect on the excitations by scaling the strength of the
electron-electron interaction. Here, however, we include the full
electron interaction and instead vary the physical length of the
nanostructure.

We find a novel range of unique, prominent mixing phenomena that become
evident as we scale the length. Recent papers documented coupled 2-D
plasmon-exciton transitions \cite{ho2018,chiu2017} in crystalline films of oriented
mixtures of sSWCNTs and mSWCNTs. The results demonstrated ``avoided
crossing'' effects as the tubes are \emph{shortened} (the plasmon
wavevector $q = 1/2L$ \emph{increases}) and the 2-D plasmon frequency is
increased, and approaches the exciton frequency. If the coupling is
strong, the crossing is avoided, leading to an exciton--plasmon
hybridization with a minimum energy separation, $\hbar\Omega$, known as the vacuum Rabi
splitting (VRS) \cite{hopfield1958}. A good fit to polariton theory and models was
obtained \cite{ho2018,chiu2017}. As mentioned in the introduction we find that the
1-D case introduces quite different effects.

For the shortest CNTs, we observe some avoided crossing effects and hybridization of the resonances as the CNTs are \emph{lengthened} in the (7,7)mSWCNT case.
 We find both similarities and differences when compared to the extensive literature on Rabi splitting and polariton states in the 2-D case  \cite{kockum2018,baranov2018}.
For the shortest tubes in the (3,3) SWCNT case, avoided crossing effects also manifest, but with even stronger hybridization.
To investigate the complex mix of strongly coupled resonances, we developed a new TDDFT simulation technique using sinusoidal excitation,
which produces unique strong harmonic generation effects. Our findings are next summarized and interpreted  in detail.

\subsection{From single particle excitations to T-L plasmons}\label{sec31}

The Tomonaga-Luttinger liquid model should apply to plasmons in some of
the armchair nano-tubes simulated here. Backscattering and Umklapp
scattering, leading to momentum transfers of the order of 1/a (where a
is the carbon-carbon distance), can be neglected for larger diameter
tubes as forward scattering processes dominate \cite{kane1997}. An
analytical expression for the Plasmon velocity was derived in
\cite{kane1997} for (N, N) mSWCNTs:
\begin{equation}
v_P=\sqrt{v_F\Big(v_F+\frac{8e^2}{\epsilon\pi\hbar}\ln{\frac{R_s}{R_t}}\Big)}.
\label{eq2}
\end{equation}

This formula, with $v_F$ being the Fermi velocity,
$R_s$ the screening length and $R_t$ the tube
radius, should be valid provided $N>10$. In Ref.~\citenum{shi2015}
it was shown that a physically intuitive screening length for the
resonant Plasmon case of $\lambda_P/(2\pi) = L/\pi$ produced an excellent
fit with the measured data. This choice also agrees with the accepted
value of the screening length for plasmons as defined by Pines and
Nozie'res (Ref.\citenum{pines1966}, page 326). We shall adopt this convention and
consequently we can rewrite Eq.~\ref{eq2} as,

\begin{equation}
\frac{1}{g}=\frac{v_P}{v_F}=\sqrt{1+\frac{8e^2}{\epsilon\pi\hbar v_F}K_0\left(\frac{\pi R_t}{L}\right)}\simeq
\sqrt{1+\frac{8e^2}{\epsilon\pi\hbar v_F}\ln\left(\frac{L}{\pi R_t}\right)},
\label{eq3}
\end{equation}
where the modified Bessel function $K_0$ can be
approximated with the natural logarithm provided the tube is
sufficiently long. We used the Bessel function version of Eq.~\ref{eq3} in all
calculations quoted below.

The lowest energy excitations develop into a plasmon which can be seen
from the GPI and charge densities in Figures~\ref{fig1} and \ref{fig2}. The phase
velocity, computed as $v_P = \lambda_Pf = 2Lf$
where $L$ is the length of the tube, is presented in Tables~\ref{tab1} and
\ref{tab2}, and
the results are discussed below in relation to Eq.~\ref{eq3}. We have
simulated some (10,10) nanotubes but their size becomes prohibitive when
increasing the number of unit cells and we have focused mainly on (7,7)
and (3,3) tubes. Here we want to emphasize that Eq.~\ref{eq3} only predicts
the ratio $v_P/v_F$. Our simulations and the s-SNOM measurements in
Ref.~\citenum{shi2015}, on the other hand, yield absolute values for $v_P$. We therefore
first compare our simulated $v_P$ values for the (3,3) case with the
experimental values from Ref.~\citenum{shi2015}. We extrapolate our simulated
plasmon velocities from $L=20nm$ (80uce) to $L=40nm$ (160uce) and find
$v_P = 3.49\cdot 10^6 m/s$. We can read an average measured $v_P = 3.12\cdot 10^6 m/s$ from
Figure 3b of Ref.~\citenum{shi2015}, for similar length tubes, assuming the value of $v_F= 0.8\cdot 10^6 m/s$ used in that reference. 
Scaling the diameter dependence from
$d=0.4nm$ for the (3,3) case to the average diameter of $1.45nm$ in
Ref.~\citenum{shi2015} we use the measured data of the diameter dependence of $v_P$
in Ref.~\citenum{wang2019},
and find $v_P = 2.9\cdot 10^6 m/s$ from our simulations, which agrees within $7\%$
with the average of the measured values in Ref.~\citenum{shi2015} ($3.12\cdot 10^6 m/s$). This
establishes the fact that our simulated plasmon velocities are
consistent with the experiments in Ref.~\citenum{shi2015} within expected error
limits.

In order to further compare our simulated plasmon velocities with
Eq.~\ref{eq3} we match the results for $v_P/v_F$ from Eq.~\ref{eq3}  to the experimental
results in Ref.~\citenum{shi2015} by adjusting the prefactor multiplying the
K0-function. This is done for the (7,7) case which has a diameter large
enough that the theory should be applicable. The same equation is then
used for the (3,3) case while adjusting for the different diameter. The
experimental results in Ref.~\citenum{shi2015} were shown to be in
agreement with $v_P/v_F$ as predicted by Eq.~\ref{eq3} using a value of $v_F=0.8\cdot 10^6 m/s$. 
We compare our simulated values for $v_P/v_F$ for the (3,3)
case in Table~\ref{tab1}. The value of $v_F$ is typically taken to be in the range
$0.8\cdot 10^6 m/s$ to $1.0\cdot 10^6 m/s$ but was found to be $0.635\cdot 10^6 m/s$ for the (3,3) CNT
from a DFT band-structure simulation \cite{li2024}. We use this value of $v_F$
in Table~\ref{tab1}. Our simulated plasmon velocities for the shorter tubes are
considerably slower than the predicted values from Eq.~\ref{eq3}, as might be
expected since these tubes have diameters that are more comparable to
the length. This should lead to back scattering and Umklapp scattering
effects that decrease the velocity. As the length increases, however,
the simulated $v_P/v_F$ approaches that of $v_P/v_F$ from Eq.~\ref{eq3}. As noted
earlier, we expect worse agreement with the theory for the N=3 case, due
to the tube's small diameter (i.e., since $N<<10$).
Nevertheless, for the longest (3,3) case simulated, 80uce, we find a
difference of only $+15.9\%$ as well as $+17.5\%$ for the extrapolated
160uce case.

For the (7,7) case the longest tube length is 5nm (20uce), and while
extrapolation to $40nm$ is not feasible, we note that the difference
between the simulated $v_P/v_F$ and the theoretical prediction of $v_P/v_F$
decreases progressively as the length is increased, see Table~\ref{tab2}. The
difference is $22.5\%$ for the longest tube (20uce). We use a value of
$v_F=0.787\cdot 10^6 m/s$ based on a DFT simulation \cite{li2024}.

\begin{table}[htbp]
  \begin{tabular}{lllllll}\hline
Number of unit cells & 5 & 10 & 20 & 40 & 80 & 160 \\ \hline 
Length(nm) & 1.247 & 2.493 & 4.994 & 9.989 & 19.978 & 39.89\\
Resonant Energy, eV & 1.62 & 1.23 & 0.825 & 0.525 & 0.335 &
0.181*\\
$v_P$ simulated ($10^6m/s$) & .98 & 1.49 & 1.98 & 2.54 & 3.17 &
3.49*\\
$v_P/v_F$** & 1.54 & 2.35 & 3.12 & 4.00 & 4.99 & 5.50*\\
$v_P/v_F$ (Eq.~\ref{eq3}) & 2.345 & 2.918 & 3.433 & 3.891 & 4.3040 &
4.681\\
Error (\%) & -34.3 & -19.5 & -9.12 & +2.8 & +15.9 &
+17.5*\\ \hline
\end{tabular}
  \caption{Comparison of simulations and theory for the (3,3)mSWCNT.\\
    *Extrapolated, ** $v_F = 0.635\cdot 10^6 m/s$}
\label{tab1}
\end{table}

\begin{table}[htbp]
  \begin{tabular}{llll}\hline
  Number of unit cells & 5 & 10 & 20 \\ \hline   
Length(nm) & 1.247 & 2.493 & 4.994\\
Resonant Energy, eV & 1.33 & 1.06 & 0.717\\
$v_P$ simulated ($10^6m/s$) & 0.802 & 1.254 &
1.71\\
$v_{P}/v_{F}$** & 1.02 & 1.59 & 2.17\\
$v_{P}/v_{F}$ by Eq.~\ref{eq3} & 1.62 & 2.21 & 2.80\\
Error (\%) & -37.0& -28.1& -22.5\\ \hline
\end{tabular}
  \caption{Comparison of simulations and theory for the (7,7)mSWCNT.\\
    ** $v_F = 0.787\cdot 10^6 m/s$}
\label{tab2}
\end{table}

We next discuss the 4-D pictures, 1-D pictures and GPI values in Figure~\ref{fig2} (the (7,7)mSWCNT case). Peak 1 for the 5uce has a velocity derived
from Eq.~\ref{eq1} close to the Fermi velocity, i.e. $g = 1$, so this is
a resonance of the type that was experimentally measured in Ref.~\citenum{zhong2008} and
also simulated for the (3,3)mSWCNT case in our earlier
paper \cite{polizzi2015}. The GPI is only 0.6 for the 5uce, in
agreement with the velocity and $g=1$. This resonance has unique
4-D and 1-D distributions not seen for the longer tubes, see Figure~\ref{fig2}.
The 4-D plot shows that there are four regions in the center of the tube
in which charges oscillate between high and low density (red and blue
colors), while the ends of the tube have wider concentrations of
opposite polarity. The latter are expected for an incipient plasmon type
resonance. This behavior is confirmed in the 1-D plot. The lowest energy
resonance grows rapidly in amplitude and sharpness, and decreases in
resonant energy, as L increases. The GPI also grows to 1.44 for 20uce,
and the 4-D and 1-D plots indicate T-L plasmon character.

For the (3,3)mSWCNT case the
lowest energy resonance grows rapidly in amplitude with L as it
sharpens. For this case we have no GPI values available but the 1-D and
4-D plots can be used as guidance for the type of resonances we see. For
the 5uce $g\sim 1$ and the 4-D and 1-D plots are
similar to what we discussed for the (7,7) case above. As L increases,
the low energy resonance decreases in energy and becomes more
plasmon-like as indicated by the 1-D plots and 4-D plots. These are very
similar from the 10uce to the 40uce (Figure~\ref{fig1}).

 There are weak
second and third harmonic resonances that grow with L. These will be
discussed later, see Figure~\ref{fig6}. 

\subsection{Avoided Crossings, Polaritons and Hybridized States: The (7,7)CNT}

The (7,7) mSWCNT was simulated at three different lengths, 5, 10 and 20
unit cells (uce). Figure~\ref{fig2} displays the simulated spectra for these as
well as GPI values for representative numbered resonances.

\begin{figure}[htbp]
    \centering
    \includegraphics[width=0.75\linewidth]{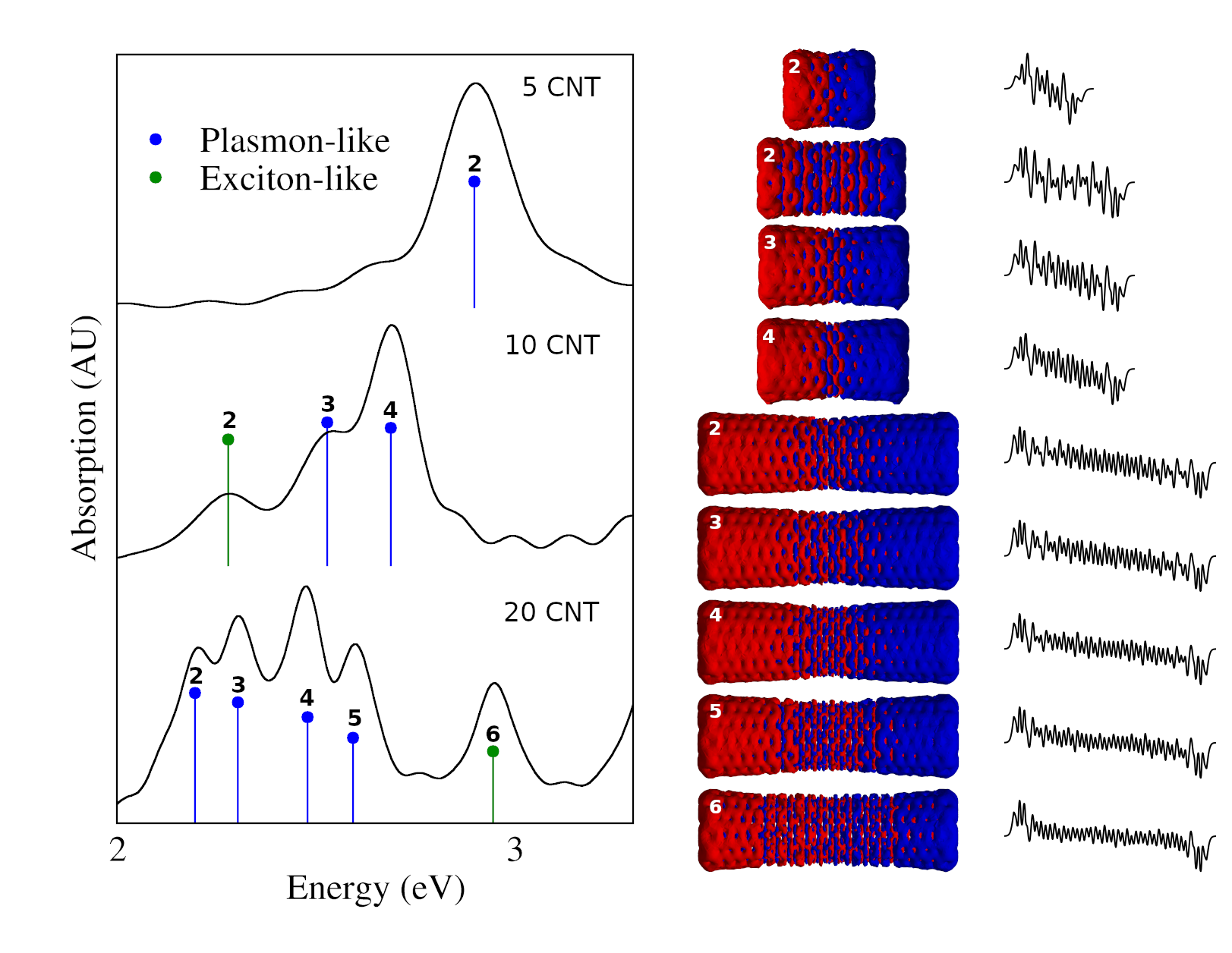}
    \caption{Spectra, 4D plots, and 1D plots for peaks from 2 eV to 3 eV
for the (7,7)mSWCNT . Top (5uce), center (10uce), bottom (20uce). GPI
values are indicated with bars. For these green color indicates b-b
character and blue color plasmon-like character.}
    \label{fig3}
\end{figure}

We discuss the peaks in the energy region from 2eV to 3.3eV. We indicate
the GPI values reported in Figure~\ref{fig2} in Figure~\ref{fig3}
with colored bars, green
for b-b and blue for plasmon-like character. For the {5uce} there
are no clear b-b transitions. Peak 2 at 2.9 eV changes to a lower energy
for the 10uce (Peak 4 at 2.69 eV, with a subpeak, \#3, at 2.5eV),
indicating its plasmon-like character{.} Peak 2 at 2.28 eV, has a 4-D
plot, and this is even more clearly seen in the 1-D plot, close to what
is expected for a b-b transition (even distribution with some spikes at
the end). In agreement with this the b-b transition for long (7,7)mSWCNT
tubes is known to be at 2.2 eV \cite{gharbavi2015}, or at 2.4 eV \cite{haroz2013}.

As L increases to the 20uce, there is a group of four split hybridized
resonances from 2.2 eV to 2.6 eV (Figure~\ref{fig3}). Peak 2 at 2.197 eV and Peak
5 at 2.595eV both have a mixed (hybridized) plasmon/b-b character. Peak
2 has a low GPI of 0.57 while GPI for Peak 5 is 0.33 (compare GPI = 1.44
for the lowest energy 0.717 eV T-L plasmon peak). There is a gap from
the four split resonances to Peak 6 at 2.947 eV which has more b-b like
1-D and 4-D pictures and a lower GPI of 0.31. This gap is interpreted as
a possible Rabi-like splitting that is due to an avoided crossing. The
lower energy four split peaks constitute the lower plasmon-type
polariton, while the 2.947 eV b-b like resonance is identified as the
upper polariton. As the length is increased from 5uce to 20uce the
polariton-like resonances switch character as expected near an avoided
crossing phenomenon \cite{forn-diaz2019}. Note that all resonances have
GPI$<1.0$, indicating that they are all hybridized, though to
different degrees.

The peaks at higher energy (peaks 7 through 16 in Figure~\ref{fig2}) show similar
effects to peaks 2 through 6 in Figure~\ref{fig3}, including an avoided crossing
(not shown).

The appearance of the four subpeaks in the lower polaritons at 2.2 eV to
2.6 eV (as well as the peaks at 3.85 eV to 4.42 eV.) is a new phenomenon
not known in previous work on polaritons or Rabi splittings. We
hypothesize that they are characteristic of strong coupling and the
ensuing hybridization in the 1-D case, which has not been explored
before.

\subsection{Strong Hybridization: The (3,3) CNT}

\begin{figure}[htbp]
    \centering
    \includegraphics[width=\linewidth]{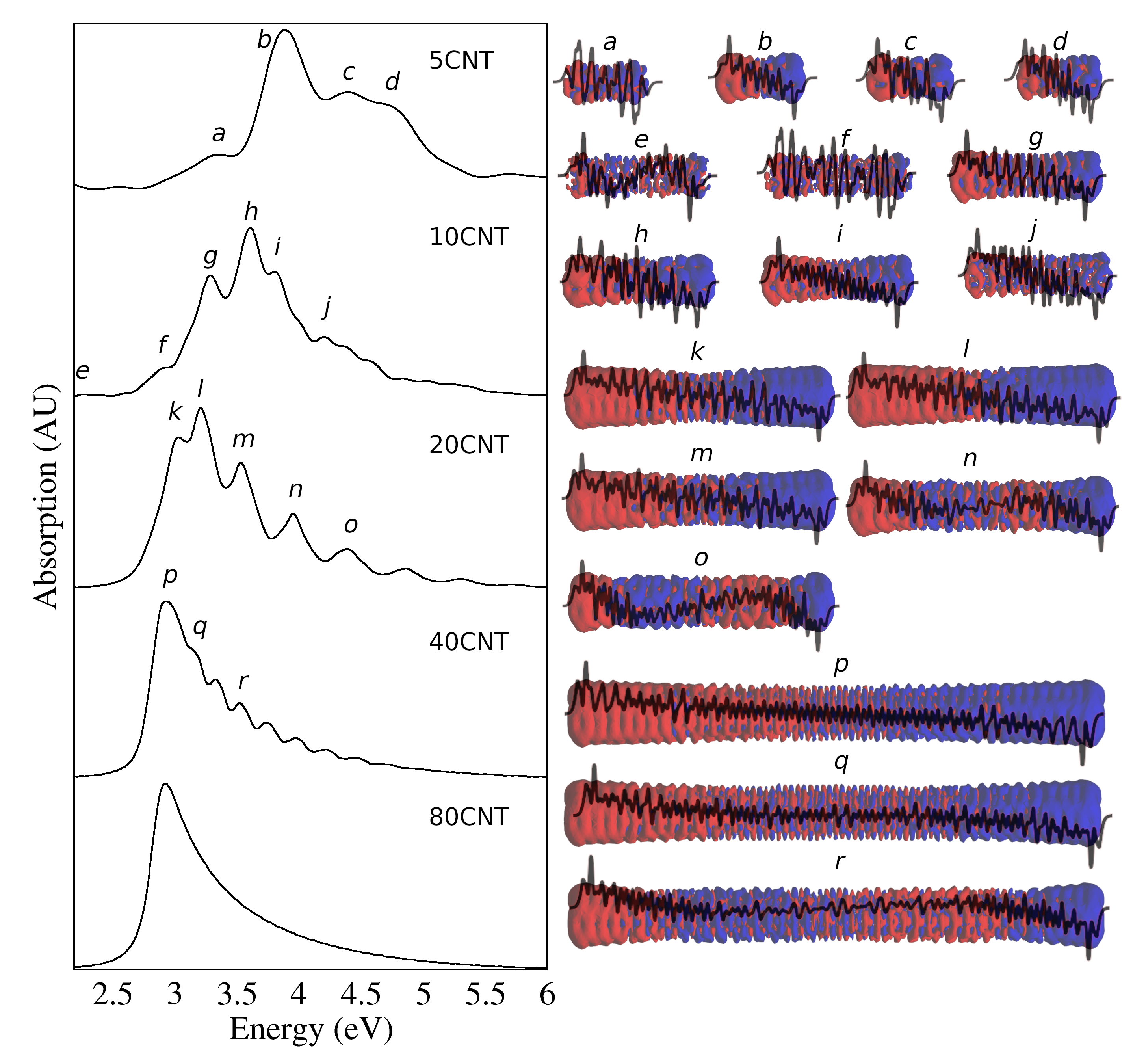}
    \caption{Spectra, 4-D and 1-D plots (black lines superimposed) for the
(3,3)mSWCNT in the energy range from 2.5 eV to 6 eV. L = 5, 10, 20, 40
and 80uce. Peaks are numbered with lower case letters.}
    \label{fig4}
\end{figure}

We summarize the development of the resonances in the energy range from
2.5eV to 6eV for the (3,3) case in Figure~\ref{fig4}. Spectra are shown to the
left and 4-D plots to the right. The 1-D plots are superimposed on the
4-D plots (black lines). Peaks are numbered with lower case letters.

\noindent \textbf{5 uce}: Peak \emph{a} coincides with the well-known b-b
transition for long tubes at 3.1eV \cite{spataru2004} and has 1-D and 4-D
excitation plots that are flat in the center of the tube, indicating b-b
character, but with some partial hybridization evident at the ends, even
for this, the shortest (3,3)mSWCNT case. Peaks \emph{b} and \emph{c} are more
plasmon-like but also hybridized. Peak \emph{d} (4.44eV) is at an energy close
to that of the $\pi$ plasmon. There is a well-known $\pi$-plasmon
\cite{kramberger2008,murakami2005}
for long tubes with an optical absorption peak energy at 4.5eV. The
$\pi$-plasmon energy should not depend on L, and our previous paper \cite{polizzi2015}
showed a peak at 4.44eV for the (3,3)mSWCNT which we identified from the
4-D plot as the $\pi$ plasmon. The $\pi$-plasmon resonance should have a more
even excitation but is partly hybridized even for the shortest
(3,3)mSWCNT.\\

\noindent \textbf{10uce}: Peak \emph{e} at 2.38eV is very weak. In the next section we identify
it as the second harmonic of the T-L resonance, compare Figure~\ref{fig6} where
it is displayed much more clearly by using a log-scale; The 3.1 eV b-b
transition (peak \emph{b} for 5uce) is hybridizing further, splitting to the
weak b-b-like Peak \emph{f} at 2.88 eV and the stronger hybridized plasmon peak
g at 3.29 eV. The 3.89 eV plasmon has moved down to two split peaks at
3.62 eV (Peak \emph{h}, the strongest peak) and 3.81 eV (peak \emph{i}), respectively.
The split of the 3.89 eV plasmon peak into two plasmons was initiated
for the 7uce (5,5) case, see Ref.~\citenum{polizzi2015}. It may be related to the
splitting of hybridized plasmon peaks into subpeaks that we find for the
20uce (four peaks) and 40uce (eight peaks) cases. The spectrum has
developed a slope toward higher energies, another feature that does not
agree with a standard Rabi splitting picture. The energy gap seen in
typical 2-D polariton cases \cite{ho2018} also never develops.
Peak \emph{j} at 4.21 eV has the character of a $\pi$-plasmon, but has almost
completely merged with the slope of the hybridized plasmons.\\

\noindent \textbf{20uce:} The peaks from 3.05 eV (k) and up to 4.4 eV (o) are
dominated by a gradually developing plasmon/b-b hybridization and a van
Hove type slope. There is no indication of the b-b peak at 2.88 eV (peak \emph{f}
for 10uce). Peaks \emph{k} and \emph{l} have more of a plasmon character
which changes toward b-b character for Peak \emph{m}. Peaks \emph{n} at 4.39 eV and
Peak o at 4.56 eV are so strongly hybridized that the 1-D and 4-D plots
have a second harmonic character, another unique feature for 1-D
hybridization. The four split peaks from \emph{l} to \emph{o} form an evenly spaced
pattern of subpeaks similar to what we found for the 20uce
(7,7)mSWCNT. The subpeaks are on top of the slowly decaying background
reminiscent of the van Hove shape characteristic of the 1-D density of
states that started for the 10uce. This sloping shape is
consistent with the conclusion in Ref.~\citenum{haroz2013} which finds
that if the exciton \emph{dominates} the spectrum one obtains a sharp
line without the long van Hove slope. The latter instead  reflects ``interaction
with the continuum'' i.e. the electron is in the conduction band, not
bound to a hole. In our case we hypothesize that the hybridization leads
to ionization of the (hybridized) excitons, and a recovery of the basic
van Hove line shape. We also note that the subpeaks are most prominent
for the 20uce case.\\

\noindent  \textbf{40uce}: The progress of the plasmon toward lower energy is now
complete and the lowest energy peak \emph{p} is no longer a b-b but a (weakly
hybridized) plasmon. The subpeaks above that in energy are more
hybridized, see Peak \emph{q} at 3.15 eV. The three highest energy subpeaks up
to 3.74 eV show some second harmonic character in the 4-D and 1-D plots,
see peak \emph{r}. The subpeaks are weaker in amplitude for the 40uce,
and roughly twice as dense, while the van Hove shape still dominates. We
conclude that the lowest resonance is now plasmon-like with some
hybridization while more b-b-like resonances develop above that, a
similar inversion in character to what we found for the (7,7) case.
There is no energy gap in the energy region from 2.94 eV to 6 eV,
however. Energy gaps are explored further in the next section.\\

\noindent \textbf{80uce}: The subpeaks disappear and the slope to the right
has an even van Hove shape. The strong  plasmon peak remains
at 2.94 eV. It is noteworthy that this plasmon peak has become
``frozen'' at 2.94eV and no longer decreases in energy as the tube is
lengthened. \\

The above survey of the development of the resonances from 2.8eV to 6eV
for the (3,3)mSWCNT case has identified many new effects not seen in 2-D
polariton cases such as for example in Ref.~\citenum{ho2018,chiu2017,gao2018}. We
explore these and other new effects further in the next section by
employing a novel method for TDDFT simulations with sinusoidal
excitation.

\subsection{Nonlinear Harmonic Generation and Coupling Effects Explored Through Sinusoidal Excitation: The 10uce (3,3)CNT.}

\begin{figure}[htbp]
    \centering
    \includegraphics[width=\linewidth]{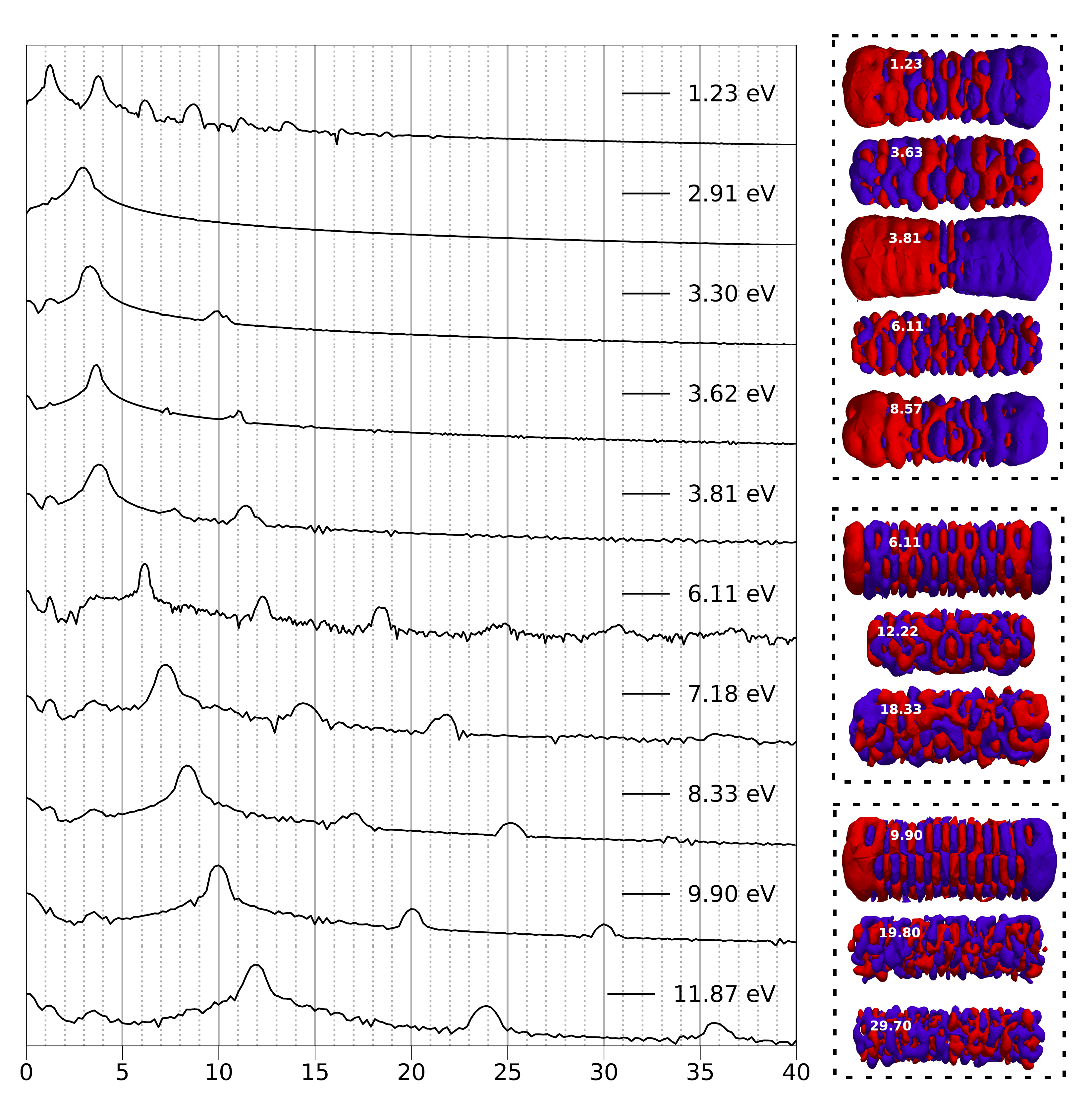}
    \caption{Spectra with sinusoidal excitation at different excitation energies.  The vertical axis uses a log
scale. The inset shows 4-D excitation plots for three selected
excitation energies, including harmonics.  Top: Excitation at the T-L plasmon energy and its odd order harmonics. Middle: Excitation at 6.11eV and its second and third harmonics. Bottom: Excitation at 9.90eV and its second and third harmonics.} 
    \label{fig5}
\end{figure}

In this section we further explore the strong hybridization effects
found for the (3,3)CNT by employing a novel method for TDDFT simulations
with sinusoidal excitation (see section on Methods).
This investigation revealed
that many resonances produce harmonic responses and also often couple
(weakly) over the entire energy range that we have simulated. We chose
the 10uce(3,3) case for this initial investigation. In this section,
sinusoidal excitations are employed to separately excite each frequency.
This method provides a clearer spectrum through Fourier transformation,
complementing the broadband excitation results obtained from fast
impulse methods. The spectra with this type of excitation are displayed
in Figure\ref{fig5}. We note that the spectra in Figures~\ref{fig1} and
\ref{fig2} were obtained
with very fast impulse excitation, which effectively applies frequencies
across the entire frequency (energy) range displayed. We will identify
this as the \emph{broadband excitation} case. The frequencies for which
the sinusoidal excitation was imposed were identified from the broadband
spectrum. The character of each response (plasmon, hybridized, b-b) is
displayed through 4-D plots in Figure~\ref{fig5}. We note that the
spectra and 4-D plots are quite different depending on whether broadband
or sinusoidal excitation is used, and will explore this below.

We can identify three main regions of resonances in the 10uce broadband
spectrum:
\begin{description}

\item[(A)] The (T-L) plasmon at 1.23eV

\item[(B)] Hybridized plasmons/excitons from 2.91eV to 3.81eV

\item[(C)] $\pi+\sigma$ plasmons at energies from 8eV to 40eV
\end{description}
  
\begin{figure}[htbp]
    \centering
    \includegraphics[width=0.9\linewidth]{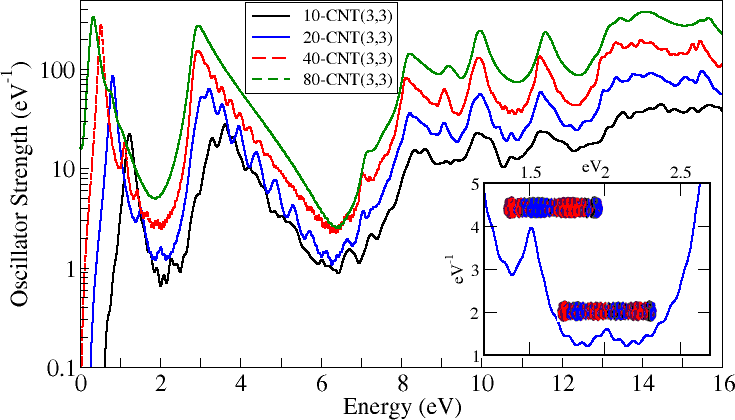}
    \caption{Broadband spectra on a log scale. The insert shows a
partial linear spectrum for the 20uce case displaying responses near the
second and third harmonics. The 4-D plots of these are also shown.}
    \label{fig6}
\end{figure}

The  $\pi+\sigma$ plasmons are ``surface'' plasmons, similar to the $\pi$ plasmon.
They have been studied extensively, first with electromagnetic theories
and later with ab initio methods, see the recent review in Ref.~\citenum{shoufie2020}.
Due to their high energies optical experiments are lacking, and instead
the spectra are measured by EELS (Electron Energy Loss Spectroscopy). An
EELS spectrum for axial excitation of single wall carbon nanotubes was
presented in Ref.~\citenum{kramberger2008}. The corresponding optical spectrum can be
deduced by extrapolating the EELS spectrum to a wavevector of 0 cm-1,
which results in a broad peak at 17.6eV.  $\pi+\sigma$ plasmons can also be
excited in the perpendicular (y) direction. There are no publications
related to  $\pi+\sigma$ plasmon experimental \emph{optical} spectra for CNTs
\emph{with} \emph{finite length}, so the data presented here in Figures
~\ref{fig1},\ref{fig2},\ref{fig5} and \ref{fig6} are unique.
While the energy ranges covered agree
approximately, the finite length spectra simulated by TDDFT show many
sharp peaks, especially as L increases, while the EELS spectra for long
tubes are smooth.

There are ``forbidden'' energy gaps at about 2eV and from about 5.5 eV
to 7.5 eV in the broadband spectrum. These energy gaps are different
from the splittings studied for 2-D polaritons \cite{ho2018,kockum2018} ,
however, since their widths do not depend on the coupling strength - as
we discussed in previous sections we interpret the coupling strength as
increasing with L whereas for the 2-D case it increases as $1/\sqrt{L}$
\cite{ho2018}. We will leave detailed exploration of the energy gap at 2 eV
for future investigation.

Figure~\ref{fig5} shows spectra for several transitions excited with sinusoidal
excitation as well as 4-D plots of the responses for three of these
cases (1.23eV, 6.11eV and 9.90eV). Using sinusoidal excitation it is now
possible to excite resonances in the upper energy gap (examples shown at
6.11eV, and 7.18eV, see Figure~\ref{fig5}), but exciting resonances outside the
gap does not produce strong responses in the gap with one exception: the
T-L plasmon at 1.23eV has a response at 6.11eV (see below). In strongly
coupled 2-D polariton cases an energy splitting is produced because the
oscillations of the two coupled resonances on either side of the Rabi
splitting gap produce electromagnetic fields out of phase with the
applied field; the fields thus cancel in the gap \cite{todorov2010,askenazi2014}. This
effect is analogous to the forbidden optical phonon (reststrahlen) band
of bulk polar semiconductors \cite{kittel1963}. For the 1-D 10uce (3,3) CNT case
we hypothesize that broadband excitation, which can excite all
frequencies simultaneously, produces responses with phases such that
they cancel, resulting in an energy gap through a process that is
similar to what occurs in the above 2-D systems. \emph{Single frequency}
excitation at frequencies in the gap does not produce such
cancellations, see Figure~\ref{fig5}.

\begin{figure}[htbp]
    \centering
    \includegraphics[width=0.9\linewidth]{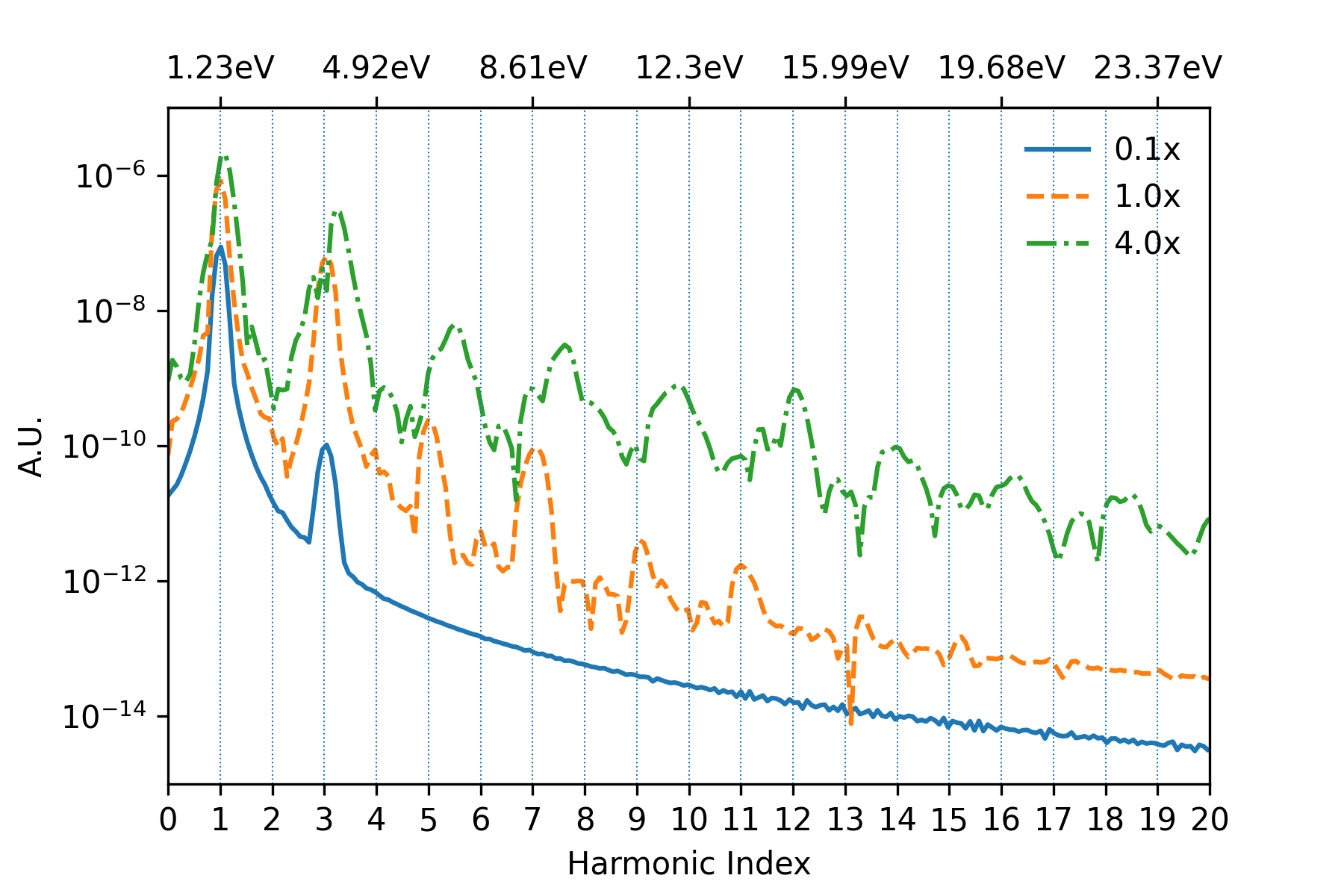}
    \caption{ Spectra with sinusoidal excitation at 1.23eV with varied amplitudes.}
    \label{fig7}
\end{figure}

An important observation is that all resonances produce \emph{harmonic}
responses. indicating that \emph{the system is nonlinear}. In order to
explore this further we performed additional simulations using
sinusoidal excitation at the \emph{T-L resonance} with excitation
amplitudes of 0.1x and 4x the amplitude we normally use, see Figure~\ref{fig7}
and Table~\ref{tab3}. For the 0.1x case only the third harmonic resulted, while
for the 1x case odd harmonics up to the eleventh were obtained. For the
4x case the harmonics are stronger, broadened, and extend further, at
least to harmonic number 17. The T-L response varies linearly with the
excitation amplitude from the $0.1\times$ case to the $1\times$ case, while it 
increased by only a factor of 3 and shifted slightly in resonance energy
for the 4x case, indicating some saturation in the latter case. This
validates our choice of displaying the spectra with the $1\times$ amplitude in
Figure~\ref{fig5}. For the 4x amplitude case, we see broadened responses as well
as a shift away from an exact harmonic frequency ratio. As the T-L
transition becomes saturated, we hypothesize that the excitation
distribution becomes more spread out compared with the plasmon type
response at lower excitation amplitudes, leading to the above effects.

We note again the unique advantage of employing sinusoidal excitation
since the harmonic resonances are very weak in the \emph{broadband
spectrum} for the 10uce case. There is a small sharp peak at 2.38eV
which corresponds roughly in energy to a second harmonic of the T-L
plasmon resonance. This peak is too weak to obtain a 4-D plot for.
Instead we show a spectrum of the low energy second and third harmonics
of the T-L plasmon with broadband excitation for the 20uce case, see the
insert in Figure~\ref{fig6}. In this case the third harmonic is much weaker than
the second harmonic, and the frequencies are slightly down-shifted with
respect to the exact second and third harmonic frequencies, whereas all
harmonics with sinusoidal excitation with 1x and 0.1x amplitude occur at
the predicted frequency. To investigate this further, we simulated the
broadband spectra of the 10uce with $0.1\times$, $1\times$ and $10\times$
amplitudes. These
spectra are all identical, indicating that the responses to broadband
radiation near the second and third harmonic frequencies are not a
non-linear effect in this case. These weak responses grow relative to
the T-L plasmon response as the length of the tube is increased. As
shown in the inset in Figure~\ref{fig6} the 4-D distributions for the 20uce case
do resemble what one would expect for (near) harmonic responses,
although as noted above they are not a nonlinear effect.

There are three resonances that show plasmonic type 4-D plots with
broadband excitation: The T-L plasmon at 1.23eV and the dipolar plasmons
at 3.62eV and 3.81eV (peaks \emph{h} and \emph{i} in Figure~\ref{fig4}). The excitation
distribution with sinusoidal excitation at 1.23eV is quite similar to
when broadband excitation is used (Figure~\ref{fig1}), indicating that this
resonance maintains its plasmon character with only minor hybridization.
Sinusoidal excitation at the T-L resonance with the 1x amplitude
produces a set of strong \emph{odd} harmonic responses as mentioned
above. 4-D plots of excitation densities for these harmonics, as well as
those of the 6.11eV and 9.90eV resonances, are displayed in the inset in
Figure~\ref{fig5}. The third harmonic of the T-L plasmon at 3.69eV is
particularly strong, down by only a factor of 12, or a conversion
efficiency of 8.3\%, when using the 1x excitation amplitude. For the
0.1x amplitude the third harmonic has a conversion efficiency to 0.0142\%,
a very strong nonlinear decrease. With this excitation amplitude the
fifth harmonic is not detectable. The third harmonic conversion
efficiencies for the 1x and 4x excitation cases are much more
comparable, 8.3\% and 14\%, respectively, whereas for the fifth harmonic
the ratio is about 10. These results are summarized in Table~\ref{tab3}.

\begin{table}[htbp]
  \begin{tabular}{llll}\hline
    Excitation amplitude & $0.1\times$ & $1\times$ & $4\times$  \\ \hline
    Fundamental & $0.1$ & $1.0$ & $3.0$\\
Third harmonic & $1.42\cdot 10^{-4}$ & $8.3\cdot 10^{-2}$ & $0.43$\\
Conversion efficiency, third harmonic & $0.0142\%$ & $8.3\%$ &
$14\%$ \\
Fifth harmonic & N/A & $2.7\cdot 10^{-4}$ &
$8.2\cdot 10^{-3}$ \\
Conversion efficiency, fifth harmonic & N/A & $0.027\%$ &
$0.27\%$ \\ \hline
\end{tabular}
  \caption{Harmonic conversion efficiencies for different excitation
amplitudes when exciting at the T-L frequency. The amplitudes are
normalized to the 1x fundamental amplitude.}
\label{tab3}
\end{table}

The third harmonic of the T-L resonance overlaps with the resonance at
3.62eV which has a 4-D plot for broadband excitation (peak \emph{h} in
Figure~\ref{fig4}) that indicates some hybridization. It couples strongly when
excited sinusoidally at the T-L plasmon frequency, producing a unique
excitation distribution (Figure~\ref{fig5} inset) with a maximum amplitude in the
center of the tube and minima at the ends. The red/blue coloring is what
would be expected for the third harmonic of 1.23eV, but with a phase
reversal. Sinusoidal excitation of the T-L plasmon yields a strongly
hybridized response in the energy gap, at 6.11eV; this corresponds to a
fifth harmonic, again with a maximum amplitude in the center of the tube
and minima at the ends, but without a phase reversal. The seventh
harmonic at 8.57eV instead has a more plasmon-like distribution. The
3.62eV and 3.81eV resonances produce weak second harmonic responses and
stronger third harmonic responses (Figure~\ref{fig5}) consistent with their
plasmonic like character with broadband excitation, see Figure~\ref{fig4}. The
resonances in the  $\pi+\sigma$ region have distributions when excited with
sinusoidal excitation (see the 9.90eV example in Figure~\ref{fig5}, inset),
typical of surface plasmons, with alternate signs for consecutive unit
cells (also compare Ref.~\citenum{kramberger2008,murakami2005}). With sinusoidal excitation at 1.23eV
the 9.90eV resonance shows a very different, more plasmon-like, distribution indicating that
the $\pi+\sigma$ plasmons are also strongly coupled to, and hybridized with,
the lower energy resonances.

Excitation at the 6.11eV resonance yields a second harmonic response at
12.22eV and a third harmonic at 18.33eV, of about equal amplitude. There
are also weak higher order harmonics. Figure~\ref{fig5} (insert) shows the 4-D
plots for representative harmonic resonances when sinusoidal excitation
at 6.11eV is employed. The 6.11eV resonance has alternating red and blue
bands indicating the potential for producing equal amplitude odd and
even harmonics, which are shown in the spectra of Figure~\ref{fig5}. The second
and third harmonic distributions are strongly hybridized and very
different from that of the $\pi+\sigma$ resonances in the same energy region.
The other resonance in the bandgap at 7.18eV, as well as the $\pi+\sigma$
resonances at 8.33eV, 9.90eV and 11.87eV in Region C, produce second and
third harmonics of about equal amplitude, as expected, since their
fundamental distributions are not plasmonic. We show the distribution
due to the fundamental sinusoidal excitation at 9.90eV which is typical
of $\pi+\sigma$ resonances. Its harmonics are again very strongly hybridized.

With sinusoidal excitation many of the resonances are coupled, i.e.
excitation at one resonant frequency produces a response at other
resonance frequencies. All higher energy resonances produce a fairly
weak response at the T-L frequency (down by three orders-of-magnitude or
more). Resonances in the energy gap and in Region C all show similar
broad, coupled, responses in region B and a peak at the T-L energy, see
Figure~\ref{fig5}. As the excitation frequency is increased coupled responses in
the lower energy regions fall off faster and the 11.87eV resonance has a
broad dip in the energy gap. As we saw previously, the resonances in
region B all couple and hybridize together. The resonances at 3.30eV and 3.81eV have the strongest
coupled responses down to the T-L energy. The b-b exciton resonance at
2.91 eV is unique in that it shows no harmonic response and no coupled response at 1.23eV.

We note that our exploration of nonlinear harmonic generation in
\emph{finite length} mSWNTs that show plasmon and polariton effects is
the first of this kind as far as we are aware. It would be of interest
to pursue further such studies in the future. High harmonic generation
in \emph{long} armchair (9,9)SWCNTs was analyzed in Ref.~\citenum{zurron-cifuentes2020}. This
paper performed calculations using sinusoidal excitation with a pulse
length of about three periods. The band structure was obtained through
tight binding methods, whereupon the response to the pulses was found by
solving the time-dependent Schr\"odinger equation. Odd harmonics were
produced up to about the thirteenth order (similar to our 1x amplitude
T-L plasmon case) for an applied intensity of $5\cdot 10^{10} W/cm^2$.
Harmonics up to the 40th order were obtained when exciting at close to
$10^{13} W/cm^2$, the damage threshold of the CNTs. The CNT in this
case behaves as a two-band system, i.e. harmonic generation relies on
what we have referred to as b-b transitions. Similar results were
reviewed in Ref.~\citenum{devega2020}. That reference simulated finite (but quite
long, 7nm) (3,3)mSWCNTs by solving the single-particle density matrix
equation of motion. It did not show evidence of plasmon effects, but
emphasized the importance of including e-e interactions in the
simulations. Ref.~\citenum{devega2020}  also did not find the broadening and
frequency shift effects that we display in Figure~\ref{fig7} for the 4x amplitude
case. The presence in our simulations of (3,3)mSWCNTs of plasmonic as
well as b-b, and hybridized versions of such resonances, clearly
indicates a much more complex series of possible harmonic outputs.

Future explorations involving sinusoidal excitation of \emph{longer}
(3,3)CNT tubes for which the coupling is stronger appear to hold promise
for better understanding of the harmonic and coupled resonance effects
we have uncovered in the 10uce case.

\section{Conclusion}

In this paper we have presented a priori TDDFT simulations of two sets
of armchair CNTs, the (3,3) and (7,7) cases. Due to advances in TDDFT
simulations, we have been able to extend the CNT length by an order of
magnitude (to 80uce or 20nm) in the (3,3) case, compared to our
previously published results \cite{polizzi2015}, and also add simulations of the
(7,7) case with a maximum length of 20uce or 5nm. The T-L plasmon
velocity for the T-L plasmon has been obtained from the shortest
``molecular'' CNTs to the longest tubes (in the (3,3) case), for which
it extrapolates in good agreement with the theory in
Ref.~\citenum{kane1997} and with the experimental results of
Ref.~\citenum{shi2015}. Plots of the excitation density in 4-D and 1-D
are presented that give a visual picture of the excitations at the
plasmon and other resonances. With further extensions of our TDDFT
techniques now underway simulations can be extended to longer lengths
and to tubes with larger diameters.

Beyond the T-L strong plasmon resonances, at higher resonance energies
we have for the first time performed a detailed exploration of 1-D
polaritons that arise  due to the hybridized coupling of single electron
(exciton) resonances and T-L or dipolar plasmons. The polaritons for the
(7,7) case exhibit an energy gap similar to the case of 2-D polaritons
with Rabi splitting. The lower energy polariton is split further into
subpeaks in a manner not found for 2-D polaritons. Such effects also
occur in the (3,3) case, but this case has a clear  energy gap at higher
energies, evident as L is increased. To explore the (3,3) case further,
we employ a novel TDDFT technique with sinusoidal excitation. This
technique produces significantly different spectra than the standard
TDDFT technique with impulse (broadband) excitation, and clarifies how
resonances couple in a complex system of resonances. We have also shown
that the nonlinearity of the system creates harmonic responses of all
resonances (except for the b-b resonance at 2.91eV). The characteristics of the CNT 1-D polaritons
are thus shown to be fundamentally different from those of the
extensively studied 2-D polaritons (hundreds of references) for which the Hopfield model \cite{hopfield1958}
has proven to match simulated and experimental data.  Further studies of 1-D polaritons would therefore be of great
interest and there are interesting objects for further studies such as the organic acene molecules
represented in our previous paper \cite{polizzi2015}.

Armchair SWCNTs can be fabricated with lengths down to 30nm or less as
demonstrated in Ref.~\citenum{wang2020}. That reference showed that
these short CNTs act as nanocavity resonators with Q-values of about 10
and with record Purcell enhancement factors. Such low loss resonators,
which we have analyzed as a function of length in this paper, can be
used to accomplish strong light matter coupling to molecules, and
facilitate many other nanophotonic applications. For example, our
results  suggest a novel set of experiments in the
near UV: The strong resonance at 2.94eV for a 80uce (3,3) CNT (L=20nm),
which we have simulated with broadband excitation, could be excited by a
laser at 421nm wavelength. The CNT resonator could be coupled to
molecules and other objects in a unique integrated circuit experiment
\cite{rossi2019,devega2016}.
Moreover, excitation at 421nm should produce outputs at odd
harmonic frequencies in the UV/EUV range, based on the fact that it arises
through resonances up to about 4eV having moved down in energy,
coalescing into the strong van Hove shape response at 2.94eV. Our
results  indicate that these types of resonances produce
strong harmonics, that would extend into the UV/EUV range. Similarly,
the T-L plasmon resonance could be employed for light matter coupling
experiments with longer tubes in the NIR/FIR range. When we extend our
simulations with sinusoidal excitation to longer CNTs, we plan to
explore such possibilities. Plasmon frequencies are of course tunable by
varying the length of the tube. In this connection we want to emphasize
that the s-SNOM experiments in Ref.~\citenum{wang2020,shi2015} employ local
excitation with a movable, sharp tip while we assume a constant electric
field parallel to the CNT. As mentioned earlier our results are
consistent with those of Ref.~\citenum{shi2015} but produce different
excitation patterns due to the different excitation methods.

In conclusion, the NESSIE modeling framework (see section on Methods) presents promising opportunities for addressing the numerical challenges in large-scale excited-state calculations within TDDFT. It not only facilitates a diverse array of electronic spectroscopy studies but also allows for in-depth exploration of nanoscopic many-body effects, including plasmonic phenomena, ranging from complex molecules to finite-size nanostructures.

\section{Methods}\label{sec2}

Spectroscopic techniques are essential tools for probing matter:
incoming radiation disturbs the sample, and the resulting response is
measured. Because this process inherently excites the system, a
first-principle modeling approach like DFT, which calculates
ground-state properties, cannot adequately capture the system's
response. TDDFT has proven highly effective in modeling how
electromagnetic fields interact with matter and in generating
spectroscopic data, including absorption and emission spectra. These
results can be quantitatively compared with experimental data, when
available, often yielding satisfying agreements.

A crucial factor enabling the new studies presented in this paper is the
significant enhancement in the numerical simulation capabilities of our
NESSIE software \cite{kestyn2020} for performing real-time TDDFT ab-initio
simulations of much longer 1-D carbon-based nanostructures, as presented
earlier \cite{polizzi2015}. NESSIE is an electronic structure code that
employs real-space finite element (FEM) discretization and domain
decomposition to carry out both all-electron (full-core potential)
ground-state DFT and real-time excited-state TDDFT calculations. It is
designed to the linear scaling capabilities of real-space mesh
techniques and leverage multi-level parallelism for targeting systems
with numerous distributed-memory compute nodes. Custom numerical
algorithms have been developed to efficiently handle the eigenvalue
problems and linear systems that are central to the software's linear
algebra operations. More particularly, the modeling approach is tailored
to optimally take advantage of the state-of-the-art FEAST eigensolver
\cite{polizzi2009}, which can achieve significant parallel scalability on modern
high-performance computing (HPC) architectures.

All our calculations were conducted using the
all-electron/DFT/TDDFT/ALDA level of theory, which has proven adequate
for accurately investigating a broad spectrum of nanoscopic many-body
effects, such as plasmonic phenomena. In addition, the CNTs were
suspended in vacuum and the effective permittivity $\epsilon$ was taken as 1.0.

To perform real-time TDDFT simulations, ground state DFT wavefunctions
are non-linearly propagated in time following a certain perturbation. At
each time step, the new electron density is calculated, enabling the
determination of the induced dipole moment as a measure of how far the
electron density has deviated from its ground-state value along a given
direction of the nanostructure (mainly along the longitudinal direction
for the 1-D structures studied in this work). The imaginary part of the
dipole's Fourier transform yields the dipole strength function, from
which the absorption spectrum is obtained, along with the expected ``true
many-body'' excited energy levels.

In real-time propagation, the system can be perturbed by applying any
external electric field. In order to capture the broadband response of
the system, it is common to use either an initial step potential or an
impulse excitation \cite{Yabana2006}.

The electron dynamics associated with a specific peak can be further
investigated by computing and visualizing the isosurface of the
frequency response density in four dimensions. These simulations aim to
provide deeper insights into the electron dynamics of particular
frequency resonances, offering relevant information about their
characteristics. The frequency response density reflects the change in
electron density resulting from an excitation at a specific resonance,
indicating charge oscillations at that resonance frequency. In practice,
after identifying the peaks and resonances of interest in the broadband
spectrum following an impulse excitation, the next step is to conduct a
new TDDFT simulation. This simulation computes the frequency response
density by performing an on-the-fly Fourier transform of the
time-varying three-dimensional electron density for all the frequencies
of interest.

To complement the broadband excitation results obtained from fast
impulse methods, sinusoidal excitations were  also employed   to
separately excite each particular frequency, allowing for a detailed
investigation of hybridization effects and revealing harmonic responses
and couplings across the entire energy range. In practice, a uniform
time-varying electric field applied along the 1-D structure is achieved
by using an oscillating potential at the target frequency. To mitigate
any potential residual broadband response from the sinusoidal signal, a
Hanning filter is applied to both the input and output signals.

\bibliography{kpy}

\end{document}